\documentstyle[preprint,aps,prl]{revtex}
\begin{document}
\bibliographystyle{prsty}
\draft
\title{
Structural, Magnetic and Electronic Properties of
Fe/Au Monatomic Multilayers}
\author{Zhu-Pei Shi, John F. Cooke and Zhenyu Zhang}
\address{Solid State Division, Oak Ridge National Lab, Oak Ridge,
TN 37831-6032}
\author{Barry M. Klein}
\address{Department of Physics, University of California, Davis, 
California 95616}
\maketitle
\begin{abstract}
An Fe/Au monatomic multilayer,
consisting of alternating single Fe and Au layers,
has been studied
by means of the self-consistent full-potential linearized augmented
plane wave method.
We show by total energy  minimization that this artificial thin film
is in the tetragonal $L1_{0}$ ordered structure with
the ratio of the interlayer spacing to the intralayer lattice constant at
0.865. In this configuration, 
the magnetic moment in each monolayer,
the spin-polarized electronic density of states,
and the corresponding band structure
are calculated.
The results are discussed
in connection with recent experiments.
\end{abstract}

\narrowtext

Magnetic multilayers have attracted considerable attention 
over the past few years~\cite{hei:93}.
A magnetic {\em monatomic\/} multilayer,
consisting of alternating single atomic layers of magnetic and nonmagnetic
elements, is the low thickness limit of a magnetic multilayer.
One particular structure out of the stacking of alternating single
atomic layers
is the tetragonal $L1_{0}$ ordered structure shown in Fig.~1. 
Some ordered alloys are known to have this phase as
their naturally occuring structure.
For example, an FePt alloy can spontaneously order
into the tetragonal $L1_{0}$ structure by 
a traditional heat treatment~\cite{zha:91,gao:95}.
More importantly, the recent development of sophisticated growth techniques
has made it possible to easily grow quality tetragonal $L1_{0}$ 
ordered structures that exist naturally~\cite{ceb:94,mit:95},
and to fabricate such materials that otherwise do
not exist in nature.
One such example is the Fe/Au $L1_{0}$ structure.
This structure does not exist naturally in the
Fe-Au phase diagram near the equiatomic composition~\cite{mas:90}, 
but can be fabricated layer-by-layer by 
molecular beam epitaxy~\cite{tak:95}.
This tetragonal $L1_{0}$ ordered FeAu material is of great 
technological interest because it 
adds a new member to a family of ferromagnets that
may have significant application potential
in magnetic recording~\cite{zha:94,kel:95}.

The magnetic properties of the artificial Fe/Au $L1_{0}$ structure
have been characterized experimentally, but to date a complete 
theoretical study of the system is still lacking. 
In one previous study, the magnetic properties
of the Fe/Au monatomic multilayer were investigated by
the layer Korringa-Kohn-Rostoker (LKKR) method~\cite{mac:90},
but not in the tetragonal $L1_{0}$ ordered structure.
In this paper, an Fe/Au in $L1_{0}$ structure
was investigated
by means of the self-consistent full-potential linearized augmented
plane wave method (FLAPW)~\cite{and:75}.
We show, by total energy  minimization, that this artificial thin film
is ferromagnetic in the tetragonal $L1_{0}$ ordered structure.
The ratio of the interlayer to the intralayer lattice constant is
0.865, and in this structure the magnetic moment of Fe is
 $2.75~\mu _{B}$.
The spin-polarized electronic density of states (DOS) and
the corresponding band structure of the $L1_{0}$ ordered FeAu
are also obtained.
We discuss these magnetic and
electronic properties in connection with the experimental results.

The first-principles FLAPW total energy method has been shown to be highly
accurate in predicting ground state structural properties of
solids ranging from insulators to metals~\cite{sin:94}.
We apply the method here to find the optimum ratio $c/a$ of 
the tetragonal $L1_{0}$ ordered FeAu.
For the Fe/Au monatomic multilayer grown on the Au(100) 
buffer layer~\cite{tak:95},
we set the in-plane lattice constant $a=a_{Au}=4.08\AA$ in our calculations.
The Brillouin zone sampling was performed using a $12^{3}$ special
k-points mesh, which yielded 126 points in the irreducible
Brillouin zone. The muffin-tin radii of 2.35 a.u. and 2.50 a.u. 
for Fe and Au respectively, 
as well as the energy cutoff of $R_{MT}K_{max}=10$ were kept fixed
in the calculations. The Ceperley-Alder local-density expression~\cite{cep:80}
for the exchange-correlation potential was used.
Self-consistency was achieved when the total energy was stable
to within 0.01 mRy/unit cell.

The FLAPW total energy as a function of volume for the tetragonal
$L1_{0}$ ordered FeAu is plotted in Fig.~2,
along with fits to the Murnaghan equation of state~\cite{mur:44}
from which one determines the equilibrium lattice constant and
bulk modulus.
The ratio of interlayer to intralayer lattice constant
in this quasi-equilibrium~\cite{note:2} phase
was found to be $c/a=0.865$, which corresponds to a $15\%$ volume contraction
with respect to the equilibrium Au volume. 
This optimum ratio ($c/a=0.865$)
is smaller than the experimental value of 0.94~\cite{tak:95}.
The equilibrium interplanar spacing  may be near the value
 expected for the hard sphere radii of the elements.  
For example, the equilibrium ratio of 0.957 for the tetragonal 
$L1_{0}$ ordered FePt~\cite{ceb:94}
is close to the value of 0.922 \cite{shi:note} estimated from the
the hard sphere radii of fcc Pt ($r_{Pt}=$1.386$\AA$)
and bcc Fe ($r_{Fe}=$1.243$\AA$).
Same argument gives rise to $c/a=$0.857 for FeAu 
($r_{Au}=$1.443$\AA$ from fcc Au),
which is very close to our calculated ratio of 0.865. 
The quality of the film as measured by  the long range order parameter $S$
is only about 0.3, compared to $S=1$ for perfectly ordered alloys.
The experimental value of $c/a$ was determined from the x-ray diffraction
which makes direct measurements of {\em correlations\/} 
between atomic positions.
Local disorder may displace some Fe and Au atoms 
from their equilibrium positions,
and may change some Fe-Au stacking to Au-Au or Fe-Fe stacking faults.
One sees that  the value of $c$ could be affected by the local disorder,
and may become larger if more Au-Au stacking faults
occur during the growth.
This might explain the discrepancy between the experimental and
calculated values.

Magnetism of transition metals is generally enhanced
by decreasing the coordination number
or by reducing the symmetry of the system~\cite{fu:85,ric:85}.
A single monolayer of Fe has lower symmetry and a smaller
 coordination number than
that of bulk bcc Fe. It has been shown that a single
monolayer of Fe has a large enhanced
magnetic moment of 3.4$\mu _{B}$~\cite{ric:85}, 
while the magnetic moment of bulk bcc Fe is only 2.2$\mu _{B}$.
The degree of symmetry and coordination number of Fe layers 
in Fe/Au $L1_{0}$ structure are between
bulk bcc Fe and single monolayer Fe. 
One expects that the magnetic moment of Fe in the $L1_{0}$ ordered FeAu
is between 2.2$\mu _{B}$ and  3.4$\mu _{B}$. This is confirmed by our
calculations. The magnetic moments of Fe and Au 
in the Fe/Au $L1_{0}$ structure
are  plotted in Fig.~3 as a function of $c/a$. The black filled circles
for Fe are calculated values; the solid line is a guide for the eye.
The open circles are calculated values for Au  with the dashed line
as a guide for the eye.
One sees that magnetic moment of Fe increases slowly from
2.47$\mu _{B}$ to 2.97$\mu _{B}$ as the ratio $c/a$ increases
from 0.74 to 1.
The trend of enhancing the magnetization in Fe layers
results from weakening the correlation between Fe layers
as their interlayer separation increases.
The Au layers in $L1_{0}$ ordered FeAu are slightly spin-polarized
by the Fe monolayer between them. 
The magnetic moment of Au decreases from
0.088$\mu _{B}$ to 0.037$\mu _{B}$ as $c/a$ increases
from 0.74 to 1; the induced magnetization is reduced as a result of
increasing the separation between Au layers and Fe layers. 
It is worth pointing out that
the small induced magnetic moment of Au indicates weak coupling 
between the Fe monolayer and the Au monolayer.

In the structure of $L1_{0}$ ordered FeAu with $c/a$=0.865,
the magnetic moment of Fe is found to be 2.75$\mu _{B}$. 
This enhanced magnetization
has been observed~\cite{tak:95}, but the experimental value of 2.5$\mu _{B}$
is less than what we calculated. If one chooses the
experimental value of 0.94 for $c/a$, the calculated magnetic moment of Fe
is 2.88$\mu _{B}$, which is much larger than the experimental value
2.5$\mu _{B}$. The existence of the local disorder in the film
may reduce the measured magnetic moment of Fe.

The electronic density of states and band structure of the
tetragonal $L1_{0}$ ordered FeAu with $c/a$=0.865
are presented in Fig. 4 and 5, respectively.
The peaks in the spin-up DOS, as shown in Fig.~4, 
are located far below the Fermi level, and their corresponding bands are 
occupied as shown in Fig.~5(a). 
However, the spin-down band in Fig.~5(b), as expected, is partially filled.
Notice that the Fermi energy is near a local minimum of the
spin-down DOS in Fig.~4. This implies that the tetragonal $L1_{0}$ ordered FeAu
is stable.
The spin-up and spin-down DOS peaks around energy of $-5eV$, 
shown in Fig.~4, are mainly from
the d-orbitals of Au atoms. One sees that they are slightly  shifted from 
each other, which gives rise to a small magnetic moment of Au 
(order of 0.06$\mu _{B}$).
The spin-up DOS peaks around energy of $-2eV$ and spin-down DOS peaks
around energy of $0.05eV$ are mainly d-orbitals of Fe. 
The large split between spin-up and spin-down DOS peaks  explains
the large Fe magnetic moment (order of 2.75$\mu _{B}$).

The MBE achieved Fe/Au monatomic multilayer~\cite{tak:95} may become
a new member of tetragonal $L1_{0}$ family of ferromagnets
which generally exhibit a strong uniaxial magnetocrystalline anisotropy.
Current members are FePt, FePd, CoPt and MnAl-base 
alloys~\cite{zha:94,kel:95}. 
A consequence of the weak coupling between Fe layers and Au layers 
as discussed above is that 
the spins of the d-electrons of the Fe layers are more 
confined to
the c-axis direction, which leads to a large c-axis magnetic anisotropy.
This strong magnetic anisotropy perpendicular to Fe and Au atomic planes
has been observed~\cite{tak:95}.

In conclusion, we have calculated 
structural, magnetic and electronic properties of the
Fe/Au monatomic multilayer system by FLAPW method.
This artificial thin film with the tetragonal $L1_{0}$ ordered structure
has the value of 0.865 for the $c/a$ ratio. 
It is found that this Fe/Au monatomic multilayer system with
a large magnetic moment of 2.75$\mu _{B}$ for Fe may belong
to the tetragonal $L1_{0}$ family of ferromagnets.

Z.P.S. would like to thank K. Takanashi and H. Fujimori for helpful discussions.
This research was supported by
Oak Ridge National Laboratory,
managed by Lockheed Martin Energy Research Corp.
for the U.S. Department of Energy under Contract
No. DE-AC05-96OR22464,
and in part by
the Oak Ridge Institute for Science and Education (Z.P.S.).
B.M.K. was supported by the University Research Fund of the University 
of California at Davis.

\pagebreak
\begin{center}
        Figure Captions
\end{center}

Fig.~1 \hspace*{0.6cm}
The tetragonal $L1_{0}$ ordered structure.

Fig.~2 \hspace*{0.6cm}
Relative total energy as a function of volume for tetragonal $L1_{0}$ ordered
FeAu. $V_{0}$ is the equilibrium volume of fcc Au corresponding to
a lattice constant of $a_{Au}=4.08\AA$.
The filled circles are FLAPW calculated data and the solid line is
a least squares fit to the Murnahan equation of state~\cite{mur:44}.
 
Fig.~3 \hspace*{0.6cm}
Magnetic moments of Fe and Au in the tetragonal $L1_{0}$ ordered FeAu
as a function of ratio $c/a$ of interlayer to intralayer lattice constants.
The black filled circles are FLAPW calculated data for Fe with 
the solid line as a guide for the eye, and
open circles calculated data for Au along with the dashed line
as a guide for the eye.

Fig.~4 \hspace*{0.6cm}
The total electronic DOS of spin-up and spin-down 
for the tetragonal $L1_{0}$ ordered FeAu.

Fig.~5\hspace*{0.6cm}
Spin-up band (a) and
spin-down band (b) for the tetragonal $L1_{0}$ ordered FeAu.
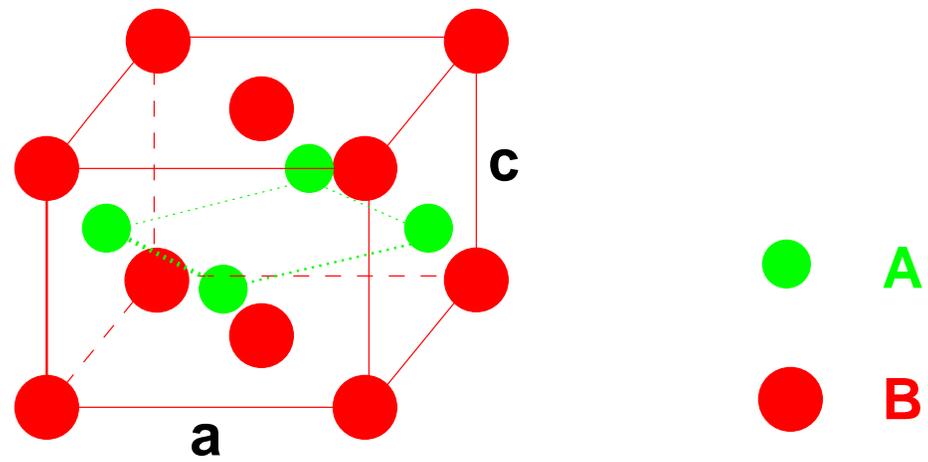
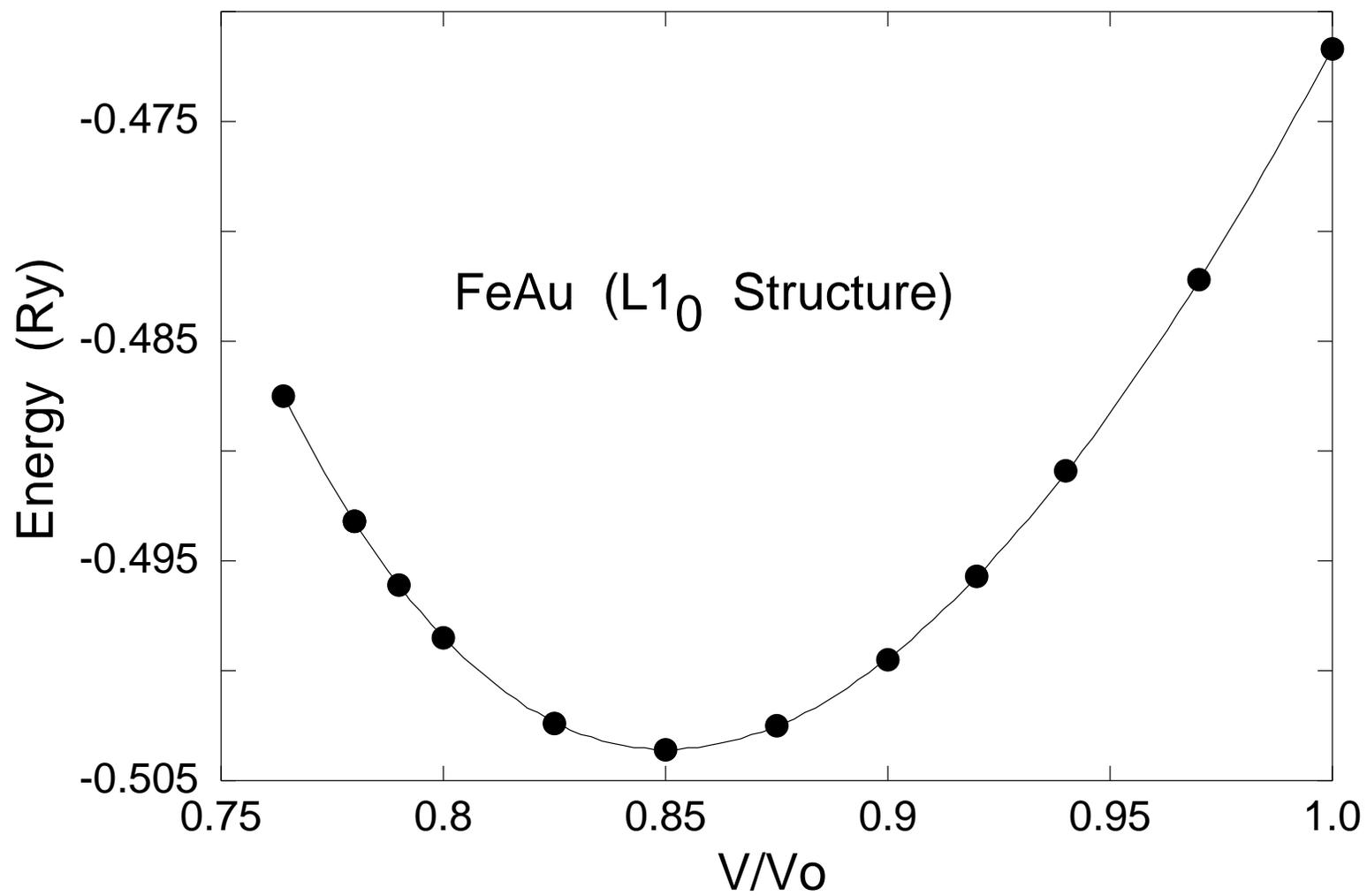
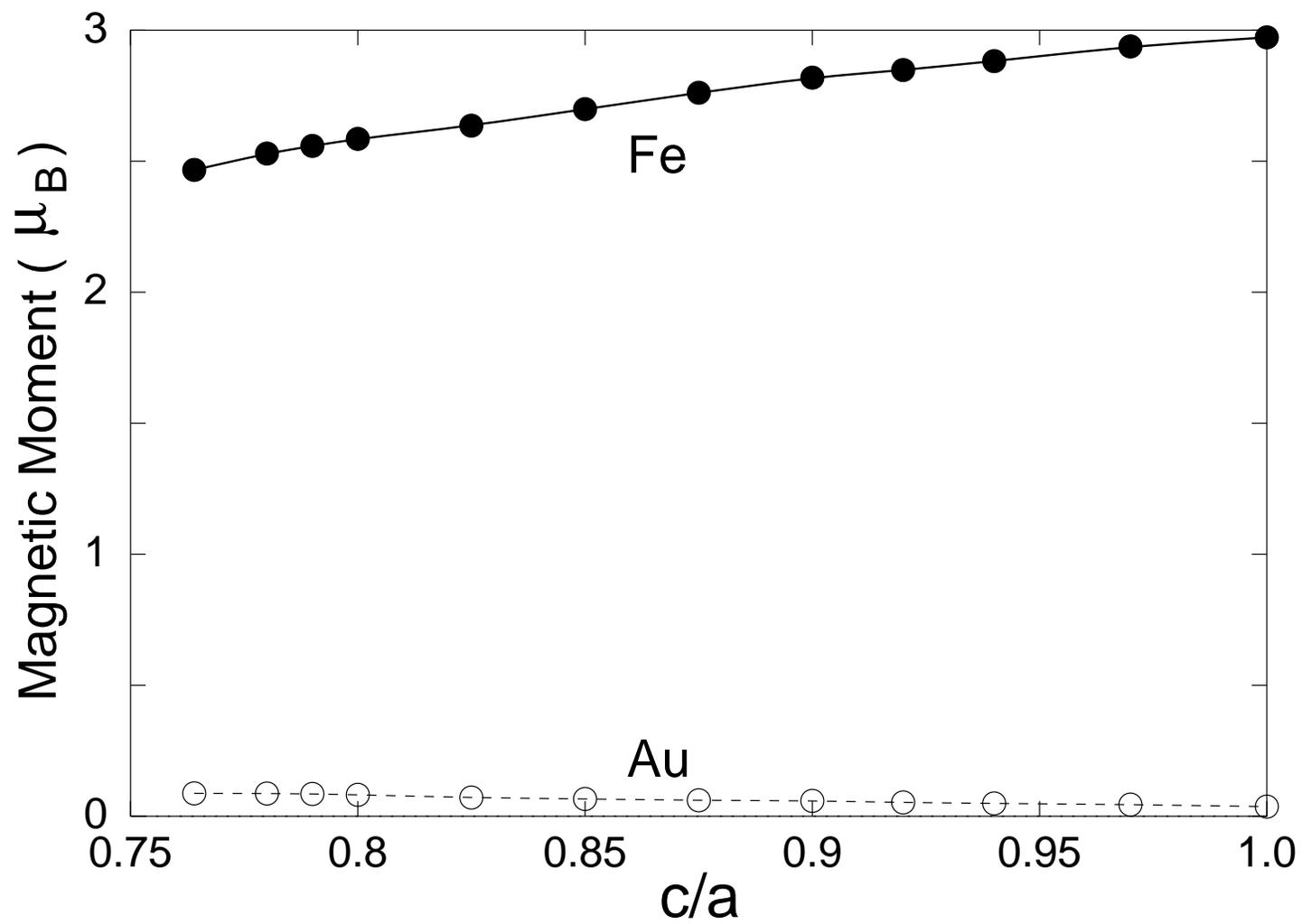
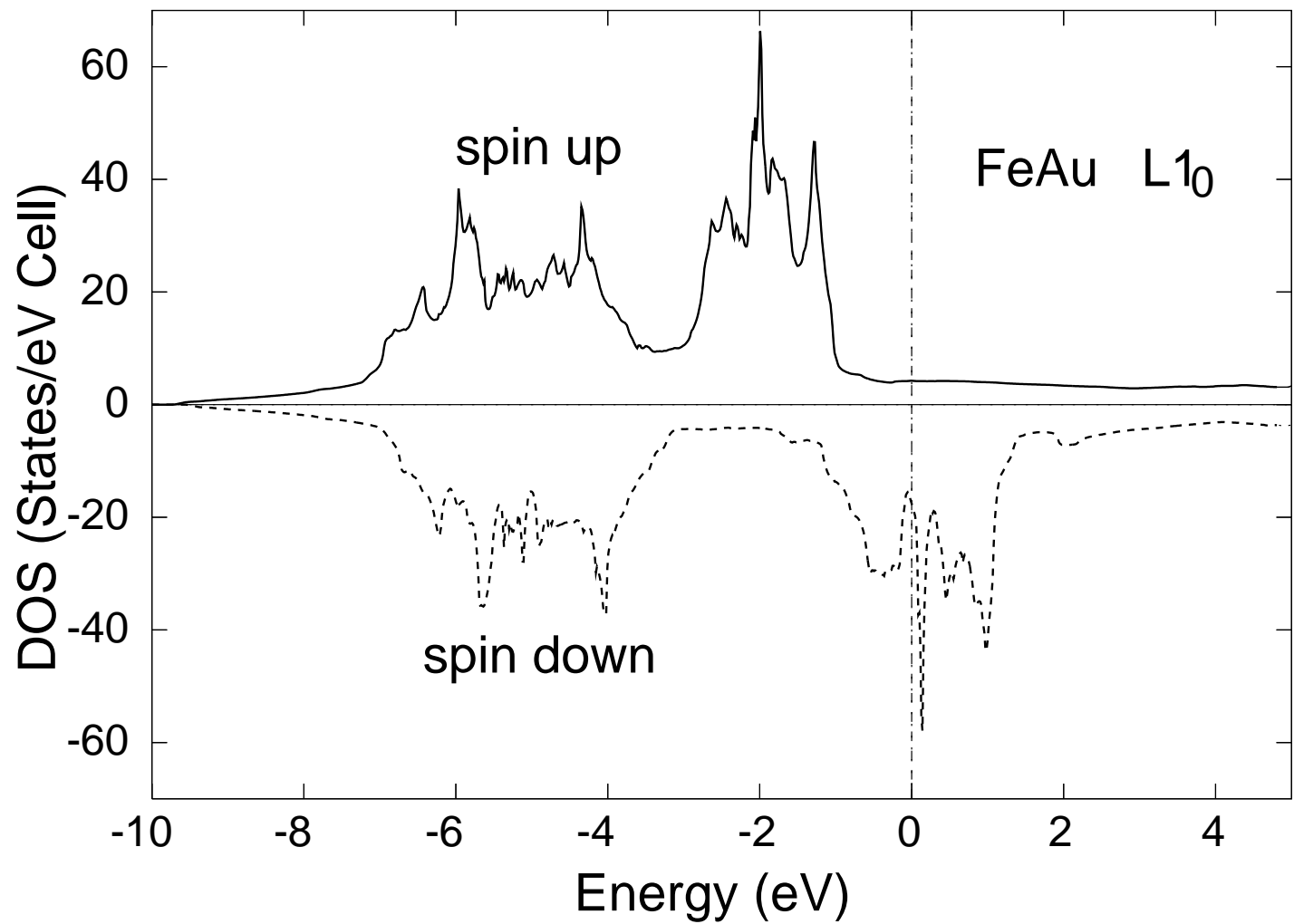
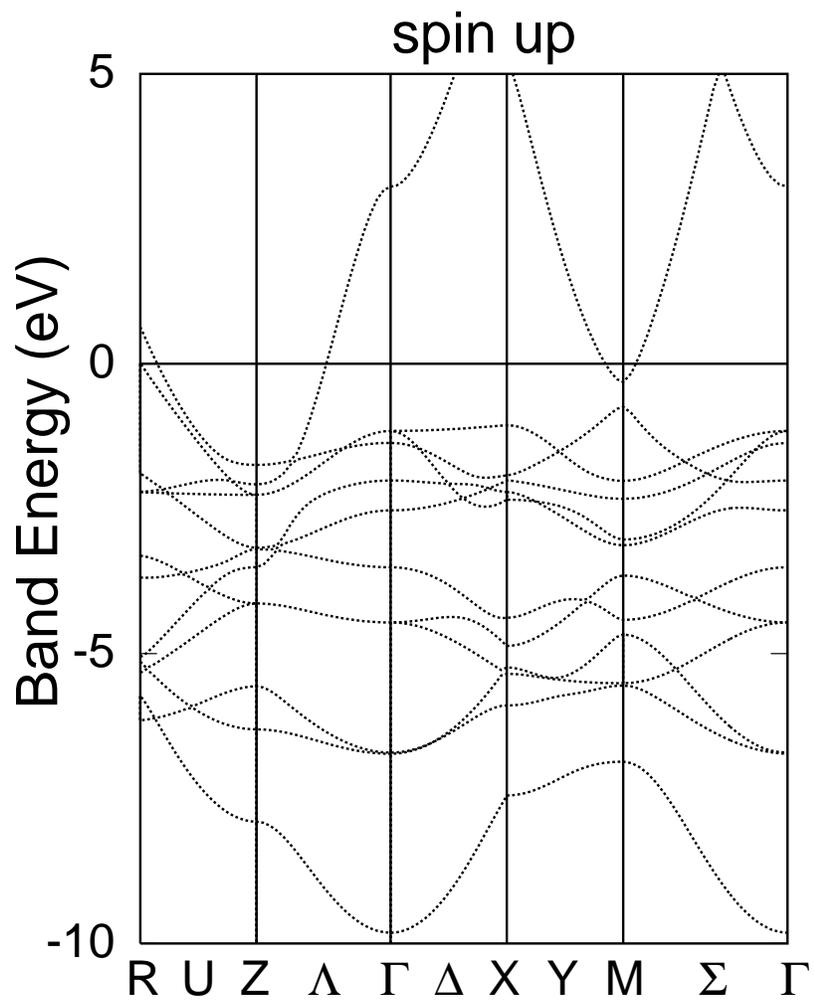
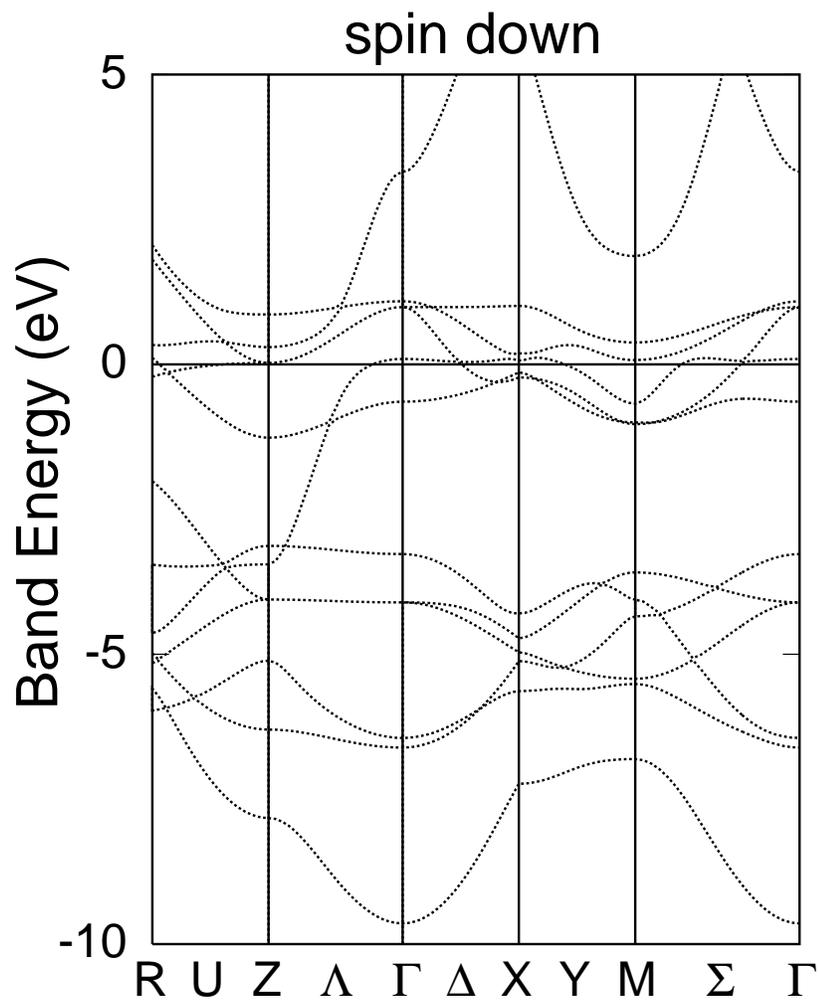
\end{document}